\documentstyle[12pt]{article}
\def\lsim{\mathrel{\rlap{\lower4pt\hbox{\hskip1pt$\sim$}}
\raise1pt\hbox{$<$}}}      %less than or approx. symbol
\def\gsim{\mathrel{\rlap{\lower4pt\hbox{\hskip1pt$\sim$}}
    \raise1pt\hbox{$>$}}}      %greater than or approx. symbol
\begin{document}
\begin{center}
{\Large \bf Explicitly broken lepton number at low  energy in the Higgs triplet model}
\vskip .3cm
\normalsize
 C. A de S. Pires   
\vskip .3cm
  Departamento de F\'{\i}sica, Universidade Federal da
Para\'{\i}ba, Caixa Postal 5008, 58051-970, Jo\~ao Pessoa - PB,
Brazil.
\vskip .3cm
\end{center}
\begin{abstract}
We suppose that lepton number is explicitly broken at low energy scale(M) in the framework of the  Higgs triplet($\Delta$) model. The scalar sector of the model is developed considering the particular assumption $M=v_\Delta \approx$ eV. We show that such assumption infers a particular mass spectrum for the scalars that compose the triplet and cause a  decoupling of these scalars  from those that compose the standard scalar doublet.  \\
%\vskip .2cm
%\noindent
%PACS numbers:\{12.60.Cn; 14.80.Mz. ; 14.60.Pq.\}\\\noindent
%Keywords:\{Lepton number violation, Higgs triplet model.\}
\end{abstract}
%%%%%%%%%%%%%%%%%%%%%%%%%%%%%%%%%%%%%%%%%%%%%%%%%%%%%%%%%%%%%%%%%
\newpage
\section{Introduction}
 Majorana mass term for the left-handed neutrinos can be easily generated in the standard model if we   enlarge its scalar sector with a scalar triplet $\Delta$ \cite{cheng-li-valle},
\begin{eqnarray} \Delta=\left(\begin{array}{cc}
 \Delta^{0} & \Delta^+/\sqrt{2} \\
 \Delta^+/\sqrt{2} &  \Delta^{++}
\end{array}
\right).
 \label{scalars}
 \end{eqnarray}
 This triplet presents the following interactions with the leptons,
\begin{eqnarray}
&&{\cal L}^Y_\Delta =	h_{ab}\overline{(\nu_{aL})^C} \nu_{bL}\Delta^0 +\frac{h_{ab}}{\sqrt{2}}\overline{\left( \nu_{aL}\right)^C}e_{bL}\Delta^+ \nonumber \\
&& +\frac{h_{ab}}{\sqrt{2}}\overline{\left( e_{aL}\right)^C}\nu_{bL}\Delta^+ + h_{ab}\overline{\left( e_{aL}\right)^C}e_{bL}\Delta^{++} +\mbox{H.c}.
	\label{leptoninteraciton}
\end{eqnarray}
In allowing $\Delta^0$ develops VEV, $v_\Delta$, automatically we have  the following Majorana mass term for the left-handed neutrinos
\begin{eqnarray}
	m_{ab}^\nu=\frac{h_{ab}}{\sqrt{2}}v_\Delta \overline{(\nu_{aL})^C}\nu_{bL}+\mbox{H.c}.
	\label{numass}
\end{eqnarray} 

In addition to the  interactions with the leptons, as showed above,  the triplet $\Delta$ also interacts with the standard gauge boson through the kinetic term 
\begin{eqnarray}
{\cal L}_{GB}=tr[\left({\cal D}_\mu \Delta \right)^{\dagger}\left( {\cal D}_\mu \Delta \right)],
\label{kinetic}
\end{eqnarray}
where
\begin{eqnarray}
{\cal D}_\mu \Delta = \partial_\mu \Delta 
-i\frac{g}{2}\left([\vec{\tau}.\vec{W}_\mu \Delta] + [\vec{\tau}.\vec{W}_\mu 
\Delta]^T \right) - i\frac{g^\prime}{2} Y \Delta B_\mu.
\label{covderiva}
\end{eqnarray}

People refer to this extension of the standard model as the  ''Higgs triplet model''. 
Depending on the mass spectrum of the scalars that compose the triplet $\Delta$, the interactions above can make  of this model a special neutrino mass model. In other words, if the mass spectrum of the scalars that compose the triplet lies in a range capable of being probed by the next generation of colliders, we  have a neutrino mass model capable of being verified  experimentally.\footnote{The phenomenology of this model was intensively  explored in various accelerators through different processes, see for example\cite{phenom1,phenom2,phenom3,phenom4,phenom5,phenom6}}.

As the triplet $\Delta$ carries two units of lepton number, thus when  $\Delta^0$ develops VEV, $v_\Delta$, lepton number is  broken, as we can see in (\ref{numass}). As is well known, in the case of a spontaneous breaking of the lepton number\cite{gelmini-roncadelli} a Goldstone boson arises, here called Majoron ($J$). However, due to  the interactions  of the triplet $\Delta$ with the standard gauge bosons given in (\ref{covderiva}), the Majoron contributes to the invisible decay  of the standard neutral gauge boson through the channel, $Z^0 \rightarrow JJJ$. It was showed in Ref. \cite{concha} that such decay rules out the scenario of spontaneous breaking of the lepton number with the triplet $\Delta$.

 The simpler solution to this problem we can envisage is the  addition of a term to the potential that breaks  explicitly the lepton number. In this case the most general gauge invariant potential we can construct with $\Delta$  and the standard doublet $\phi$, that violates explicitly the lepton number, is the following 
\begin{eqnarray}
V(\phi,\Delta)&=&\mu_\phi^2 \phi^{\dagger} \phi +  
+ \mu_\Delta^2tr(\Delta^{\dagger} \Delta) +\lambda_1(\phi^{\dagger} \phi)^2+\nonumber \\
&& \lambda_2[tr(\Delta^{\dagger} \Delta)]^2 
+ \lambda_3 \phi^{\dagger} \phi tr(\Delta^{\dagger} \Delta)+\nonumber \\
&& \lambda_4 
tr[(\Delta^{\dagger} \Delta)^2]+ \lambda_5 
(\phi^{\dagger}\Delta^{\dagger} \Delta \phi)\nonumber \\
&&-M\phi^T \Delta \phi +\mbox{H.c},
\label{potential}
\end{eqnarray}
Note that only the last term in this potential violates the lepton number explicitly.

In order to check that this solution really works, we have to develop such potential, which means to obtain the physical scalars and their respective masses. For this we have, firstly, to expand the fields $\phi^0$  and $\Delta^0$ around their VEVs 
\begin{eqnarray}
&&\phi^0 =\frac{v_\phi+R_\phi + iI_\phi}{\sqrt{2}}\nonumber \\
&&\Delta^0 =\frac{v_\Delta+R_\Delta + iI_\Delta}{\sqrt{2}},
\label{VEV}
\end{eqnarray}
and substitute these expansions in the potential above. On doing that, we obtain the  minimal condition equations
\begin{eqnarray}
&&\mu^2_\phi+\lambda_1v^2_\phi+\frac{\lambda_3+\lambda_4}{2}v^2_\Delta-\sqrt{2}Mv_\Delta=0,\nonumber \\
&&\mu^2_\Delta+(\lambda_2+\lambda_4)v^2_\Delta+\frac{\lambda_3+\lambda_5}{2}v^2_\phi-\frac{M}{\sqrt{2}}\frac{v^2_\phi}{v_\Delta}=0.
\label{constraint}	
\end{eqnarray}

This set of equations infer the following form for the mass matrix of the CP-even scalars in the basis $(R_\phi\,,\,R_\Delta)$
\begin{eqnarray}
M^2_R=\left[ \begin {array}{cc} \lambda_1 v^2_\phi& \frac{\left( \lambda_3+\lambda_5 \right)}{2} v_\Delta\,v_\phi-\frac{M\,v_\phi}{\,\sqrt {2}}
\\\noalign{\medskip}\frac{\left( \lambda_3+\lambda_5 \right)}{2} v_\Delta\,v_\phi-\frac{M\,v_\phi}{\,\sqrt {2}}& \left( \lambda_2+\lambda_4 \right) {{
v_\Delta}}^{2}+{\frac {M{{v_\phi}}^{2}}{{2\,\sqrt {2}\,v_\Delta}}}
\end {array} \right]. 
\label{MRCPeven}	
\end{eqnarray}

For  the CP-odd scalar, we obtain the following mass matrix in the basis $(I_\phi\,,\,I_\Delta)$
\begin{eqnarray}
	M^2_I= \left[ \begin {array}{cc} \frac{Mv_\Delta}{\sqrt{2}}&-\frac{Mv_\phi}{\,
2\sqrt {2}}\\\noalign{\medskip}-\frac{Mv_\phi}{\,
2\sqrt {2}}&{\frac {M
{{v_\phi}}^{2}}{4\sqrt {2}{v_\Delta}}}\end {array} \right]. 
\label{MICPodd}
\end{eqnarray}
The mass matrix for the singly-charged scalars in the basis $(\phi^+\,,\,\Delta^+)$ takes the form
\begin{eqnarray}
M^2_+=\left[ \begin {array}{cc} \sqrt {2}Mv_\Delta-\frac{\lambda_5\,{{v_\Delta}}^{2}}{2}&\frac{{\lambda_5}\,{v_\Delta}\,{v_\phi}}{2\sqrt {2}}-M{v_\phi}
\\\noalign{\medskip}\frac{{\lambda_5}\,{v_\Delta}\,{v_\phi}}{2\sqrt {2}}-M{v_\phi}&-\frac{\lambda_5
\,{{v_\phi}}^{2}}{4}+{\frac {M{{v_\phi}}^{2}}{2\sqrt {2}{v_\Delta}}}\end {array} \right], 
\label{M+matrix}	
\end{eqnarray}
while the mass of the doubly-charged scalar  takes the form
\begin{eqnarray}
m^2_{\Delta^{++}}=-\lambda_4v^2_\Delta-\frac{\lambda_5}{2}v^2_\phi+\frac{Mv^2_\phi}{\sqrt{2}v_\Delta}.
\label{delta++}
\end{eqnarray}

The factor $M$ in the potential above  modulates the size of the explicit violation of the lepton number.  If it is take large, then lepton number is violated at high energy. This  leads to the well-known type II see-saw mechanism\cite{typeIIseesaw1,typeIIseesaw2,typeIIseesaw3,typeIIseesaw4}. This mechanism  arises from the assumption
\begin{eqnarray}
	M=\mu_\Delta>>v_\phi,
	\label{highmainassumption}
\end{eqnarray}
which, when substituted in the second equation in (\ref{constraint}), gives
\begin{eqnarray}
v_\Delta = \frac{v^2_\phi}{\sqrt{2}M}.
	\label{seesawII}
\end{eqnarray}
For $M=10^{14}$GeV  and $v_\phi= 247$GeV, we obtain $v_\Delta \approx$ eV, which is the energy scale of the neutrino masses.

Phenomenologically speaking, this scenario is not so interesting because the scalars that compose the triplet $\Delta$ get very heavy and then can not be detected directly in the next colliders. However, from the theoretical side, this scenario is interesting because the smallness of the neutrino mass is being related to a new physics at high energy scale. 

Another possibility is to consider $M$ at the proper neutrino mass scale. In other word, to take $M$ at the eV scale. This possibility has received few attention in the literature. In this work we consider such possibility and then study its consequences in the scalar sector.  
%%%%%%%%%%%%%%%%%%%%%%%%%%%%%%%%%%%%%%%%%%%%%%%%%%%%%%%%%%%%%%%%%%%%%%%%%%%%%%%%%%%%%%
\section{Explicitly broken lepton number at eV scale}
The idea here is to consider that lepton number is explicitly violated at very low energy scale. We formulate this idea  through the assumed relation
\begin{eqnarray}
	M=v_\Delta<<v_\phi.
	\label{lowmainassumption}
\end{eqnarray}

We have to take care with this assumption because in the case $M=0$ we have the spontaneous breaking of the lepton number and then the Majoron emerges. Thus it is natural to expect that the case $M\neq 0$ leads to a massive Majoron with mass around $M$. Consequently, for $M \approx$ eV it should result in a very light Majoron. That would be  a problem for such assumption. Fortunately this does not happen here. To see this, note that from the two equations in (\ref{constraint}) we obtain the relation\footnote{We thanks an anonymous referee for having called my attention to this relation.}
\begin{eqnarray}
\frac{M}{v_\Delta}=\frac{\lambda_3+ \lambda_5 }{\sqrt{2}}-\sqrt{2}\lambda_1  \frac{\mu^2_\Delta}{\mu^2_\phi}.
\label{Mrelation}
\end{eqnarray}
Thus the assumption $M=v_\Delta $  is consistent with  requiring 
\begin{eqnarray}
\lambda_3=-\lambda_5\,\,\,\,\,,\,\,\,\,\,	\mu^2_\Delta=-\frac{\mu^2_\phi}{\sqrt{2}\lambda_1}.
\label{newcostraint}
\end{eqnarray}
As we will see below in (\ref{cpevenmass}) , it is $\lambda_1$ that modulates the mass of the standard Higgs. For a standard Higgs of mass about $115$GeV, this requires  $\lambda_1$  of order of $10^{-1}$. With this, from the second 	relation in (\ref{newcostraint}), we can conclude that the  scalars that compose $\Delta$, in particular the Majoron, will develop mass at the electroweak scale\footnote{This  was first observed in the Reference \cite{masiero}} . This is a nice surprising result because not only  save the scenario as turns it  phenomenologically capable of being probed in the next colliders. 

In view of this, it turns out  important to find the mass  scalar spectrum  that result from the diagonalization of the mass matrices above in the light of the  assumption made in (\ref{lowmainassumption})\footnote{
Explicitly broken lepton number with this trilinear term was first considered in the literature in the References \cite{other1}. For a more recent approach, see \cite{other2}}.

We start this task considering, first, the  CP-even Higgs. The assumption  in Eq. (\ref{lowmainassumption}) together with the first relation in (\ref{newcostraint}) leaves the mass matrix  in Eq. (\ref{MRCPeven}) with the form
\begin{eqnarray}
M^2_R=\left[ \begin {array}{cc} {\lambda_1}& \frac{M}{{\sqrt{2}v_\phi}}
\\\noalign{\medskip}\frac{M}{{\sqrt{2}v_\phi}}& \frac{1}{2\sqrt {2}}\end {array} \right]{{ v_\phi}}^{2}. 
\label{rphimatrix}	
\end{eqnarray}
First of all, note that the assumptions (\ref{lowmainassumption}) and  (\ref{newcostraint}) implies in a decoupling among $R_\phi$  and $R_\Delta$. In other word, the tiny off-diagonal elements of this matrix imply that the eigenvectors of this matrix are $H=R_\phi$  and $H^{\prime}=R_\Delta$ with the respective eigenvalues,
\begin{eqnarray}
	m^2_{H}=\lambda_1 v^2_\phi\,\,\,,\,\,\,m^2_{H^{\prime}}=\frac{v^2_\phi}{2\sqrt{2}},
	\label{cpevenmass}
\end{eqnarray}

We soon recognize $H$ as the standard Higgs.  As for the mass of $H^{\prime}$, because of the assumptions (\ref{lowmainassumption}) and (\ref{newcostraint}), it is given by
\begin{eqnarray}
	m_{H^{\prime}}= 148\mbox{GeV}.
\end{eqnarray}

Let us now consider the CP-odd scalars. The assumptions  (\ref{lowmainassumption}) and (\ref{newcostraint}) allows writting $M^2_I$  given above  in the form
\begin{eqnarray}
M^2_I=	\left[ \begin {array}{cc} \frac{{M}^{2}}{\sqrt {2}v^2_\phi} &-\frac{M}{2\sqrt {2}v_\phi}\\\noalign{\medskip}-\frac{M}{2\sqrt {2}v_\phi}&\frac{1}{4\sqrt{2}}\end {array} \right]v^2_\phi.
\label{cpoodmassmatrix}
\end{eqnarray}
which also  implies in a decoupling among $I_\phi$  and $I_\Delta$ which means that the eigenvectors  are $G=I_\phi$  and $J=I_\Delta$ with the respective eigenvalues,  
 \begin{eqnarray}
m^2_{G}=0\,\,\,\,\,\,\mbox{and}\,\,\,\,\,\, m^2_{J}=\frac{v^2_\phi}{4\sqrt{2}}.
\label{massgoldstone}
\end{eqnarray}
where we recognize that $G$ is the Goldstone to be eaten by $Z^0$, and $J$ is a massive non-standard CP-odd scalar with the following mass  value
\begin{eqnarray}
m_J= 104\mbox{GeV}.	
\label{mjprediction}
\end{eqnarray}

Let us now consider  the singly-charged scalars. The assumptions in (\ref{lowmainassumption}) and (\ref{newcostraint}) leave $M^2_+$ given in (\ref{M+matrix})  with the form
\begin{eqnarray}
M^2_{+}\cong \left[ \begin {array}{cc} \left( \sqrt{2}-\frac{\lambda_5}{2} \right)\frac{M^2}{v^2_\phi}&-\left(1+\frac{\lambda_5}{2\sqrt{2}} \right)\frac{M}{v_\phi}\\\noalign{\medskip}-\left(1+\frac{\lambda_5}{2\sqrt{2}} \right)\frac{M}{v_\phi}&\left( \frac{1}{\sqrt{2}}-\frac{\lambda_5}{4} \right) \end {array} \right] 
 v^2_\phi.
\label{massmatrixcharged}
\end{eqnarray}
Even here we are going to have a decoupling among $\phi^+$  and $\Delta^+$. As consequence  the eigenvectors are $h^+=\phi^+$  and $h^{\prime +}=\Delta^+$ with the respective eigenvalues 
\begin{eqnarray}
	m^2_{h^+}=0\,\,\,\, \mbox{and}\,\,\,\,m^2_{h^{\prime +}}=\left( \frac{1}{\sqrt{2}}-\frac{\lambda_5}{4} \right)v^2_\phi,
	\label{singlychargedmass}
\end{eqnarray}
where $h^+$ is the Goldstone that will be eaten by the standard gauge bosons $W^{+}$ and $h^{\prime +}$ is a massive singly-charged scalar with mass lying in the range 
\begin{eqnarray}
167\mbox{GeV}\leq m_{h^{\prime +}}\leq 242\mbox{GeV},	
\end{eqnarray}
for $-1\leq\lambda_5 \leq 1$.

With the doubly-charged scalar, the assumptions (\ref{lowmainassumption}) and (\ref{newcostraint}) imply
\begin{eqnarray}
	m^2_{\Delta^{++}}=\left( \frac{1}{\sqrt{2}}-\frac{\lambda_5}{2} \right) v^2_\phi.
	\label{mdelta++}
\end{eqnarray}
which gives
\begin{eqnarray}
112\mbox{GeV}\leq m_{\Delta^{++}}\leq 271\mbox{GeV},	
\end{eqnarray}
for $-1\leq\lambda_5 \leq 1$.
\section{Conclusions}
In this work we developed the scalar sector of the standard model enlarged with a scalar triplet $\Delta$ in a scenario where the lepton number is explicitly broken. Our results are based in the supposition that lepton number is explicitly broken at low energy scale ( we are thinking $M$  about eV scale) together with an assumed relation among $M$ and $v_\Delta$.

Our first result is that the relation among $M$ and $v_\Delta$  suffices to determine  the masses of the neutral scalars that compose the triplet. In the particular case $M=v_\Delta$ the model predicts  $m_H= 148$GeV and $m_J=104$GeV. For the charged scalars, the model predicts a range of value for their masses. For the singly-charged scalar we found the range $167\mbox{GeV}\leq m_{h^{\prime +}}\leq 242\mbox{GeV}$ while for  the doubly-charged scalar we found the range $112\mbox{GeV}\leq m_{\Delta^{++}}\leq 271\mbox{GeV}$.

Our second result is that,  although the  scalars that compose the triplet develop masses in the electroweak scale, they do not mix with the scalars that compose the doublet. This is a surprising and interesting result because it leaves to a clean phenomenology.

We are aware that behind such predictions there are two suppositions. At least for one of them, that lepton number is explicitly violated at low energy scale, there is explanation in the literature. Tiny $M$ can be  achieved through the shining mechanism developed in Ref. \cite{ma1,ma2}. In regard to the second supposition,  we do not know any mechanism or symmetry that determine a relation among  $M$ and $v_\Delta$.  Any idea here is very welcome.

\vskip .3cm
%%%%%%%%%%%%%%%%%%%%%%%%%%%%%%%%%%%%%%%%%%%%%%%%%%%%%%%%%%%%%%%%%%%%%%%%%%%%%%%%%%%%%
{\bf Acknowledgments.}
We thanks Alex G. Dias for useful discussion. This work was supported by Conselho Nacional de Pesquisa e Desenvolvimento - CNPq.

%%%%%%%%%%%%%%%%%%%%%%%%%%%%%%%%%%%%%%%%%%%%%%%%%%%%%%%%%%%%%%%%%%%%%%%%%%%%%%%%%%%%%%%%%%%

\end{document}